\documentclass[english,aps,superscriptaddress,preprintnumbers,reprint,footinbib,amsmath,amssymb,prb]{revtex4-2}
\usepackage[latin9]{inputenc}
\usepackage{color}
\usepackage{amsmath}
\usepackage{amssymb}
\usepackage{graphicx}
\usepackage{esint}

\makeatletter

\@ifundefined{textcolor}{}{%
 \definecolor{BLACK}{gray}{0}
 \definecolor{WHITE}{gray}{1}
 \definecolor{RED}{rgb}{1,0,0}
 \definecolor{GREEN}{rgb}{0,1,0}
 \definecolor{BLUE}{rgb}{0,0,1}
 \definecolor{CYAN}{cmyk}{1,0,0,0}
 \definecolor{MAGENTA}{cmyk}{0,1,0,0}
 \definecolor{YELLOW}{cmyk}{0,0,1,0}
}

\usepackage{aecompl}

\usepackage{epsfig}\usepackage{dcolumn}\usepackage{bm}
\usepackage{babel}

\usepackage{hyperref}
\hypersetup{
    colorlinks=true,
    linkcolor=blue,
    urlcolor=blue,
    citecolor=blue
        }

\makeatother

\usepackage{babel}
\begin{document}
\title{Analogue of atomic collapse for adatoms on rhombohedral multilayer
graphene}
\author{W.C. Silva}
\email[corresponding author:]{willian.carvalho@unesp.br}

\affiliation{São Paulo State University (Unesp), School of Engineering, Department
of Physics and Chemistry, 15385-000, Ilha Solteira-SP, Brazil}
\author{J.E. Sanches}
\affiliation{São Paulo State University (Unesp), School of Engineering, Department
of Physics and Chemistry, 15385-000, Ilha Solteira-SP, Brazil}
\author{A.M. Freitas}
\affiliation{São Paulo State University (Unesp), School of Engineering, Department
of Physics and Chemistry, 15385-000, Ilha Solteira-SP, Brazil}
\author{L.T. Lustosa}
\affiliation{São Paulo State University (Unesp), School of Engineering, Department
of Physics and Chemistry, 15385-000, Ilha Solteira-SP, Brazil}
\author{M. de Souza}
\affiliation{São Paulo State University (Unesp), IGCE, Department of Physics, 13506-970,
Rio Claro-SP, Brazil}
\author{A.C. Seridonio}
\email[corresponding author:]{antonio.seridonio@unesp.br}

\affiliation{São Paulo State University (Unesp), School of Engineering, Department
of Physics and Chemistry, 15385-000, Ilha Solteira-SP, Brazil}
\begin{abstract}
We propose that a multi-graphene of ABC-type stacking yields virtual
bound states lying within the Coulomb insulating gap of an Anderson-like
adatom. Wondrously, a virtual state constitutes the counterpart of
the atomic collapse phenomenon proposed in relativistic atomic Physics,
while the second emerges as its particle-hole symmetric, analogous
to a positron state. Thus, we introduce the effect as the adatomic
collapse, which occurs due to a flat band with a dispersionless state
and a divergent density of states $\sim|\varepsilon-\varepsilon_{F}|^{2/J-1}$
near the Fermi energy $\varepsilon_{F}$ for $J\geq3,$ where $J\pi$
is the Berry phase. We conclude this scenario based on the Kramers-Kronig
transformation of the quasiparticle broadening, from where we observe
that the aforementioned van Hove singularity induces virtual bound
states. Counterintuitively, near the singularity, we find these states
above and below the Fermi energy correlated to the existence of the
bottom and top edges of the Coulomb insulating region, respectively.
As such a behavior rises without a twist, the system is known as Moiréless
and the phenomenon emerges also assisted by the adatom Coulomb correlations.
{Similarly to Science 340, 734 (2013) we find the
effective critical atomic number $\mathcal{Z}_{c}\sim0.96$ in contrast
to an ultra--heavy nucleus. }Thus, we point out that multi-graphene
is a proper playground for testing a predicted phenomenon of the relativistic
atomic Physics in the domain of the condensed matter Physics.
\end{abstract}
\maketitle

\section{Introduction }

The atomic collapse\cite{SHYTOV2009,JR1977} is an effect predicted
by the relativistic atomic Physics, which consists of a phenomenon
wherein the charge of a nucleus overpasses its threshold and becomes
capable of catching an electron followed by a positron emission. To
observe this phenomenon ultra-heavy nuclei are required, but they
do not exist in nature spontaneously. Unfortunately, despite the efforts
in fabricating the right conditions to emulate these atoms in the
experimental native framework of this research field, the observation
of the atomic collapse remains elusive\cite{Schweppe1983,Cowan1985}.

However, some condensed matter platforms rise as successful emulators
for this effect wherein the graphene-based systems assume highly the
status to this end. In such materials, the aforementioned features
are due to the well-established low supercriticality, which can be
manufactured artificially\cite{PMID_Wang,CaIons2,Mao2016,Jiang2017}.
As examples we cite calcium ions\cite{PMID_Wang,CaIons2}, charged
vacancies and electrostatic potentials\cite{Mao2016,Jiang2017}, which
once added to graphene monolayer samples play the role of effective
supercritical atoms. As aftermath, these atoms within this supercritical
regime exhibit a localized virtual bound state appearing in the local
density of states (LDOS) as a resonant state below the Fermi level\cite{PMID_Wang}
and counterpart of the atomic collapse state.
\begin{figure}[!]
\centering\includegraphics[width=1\columnwidth]{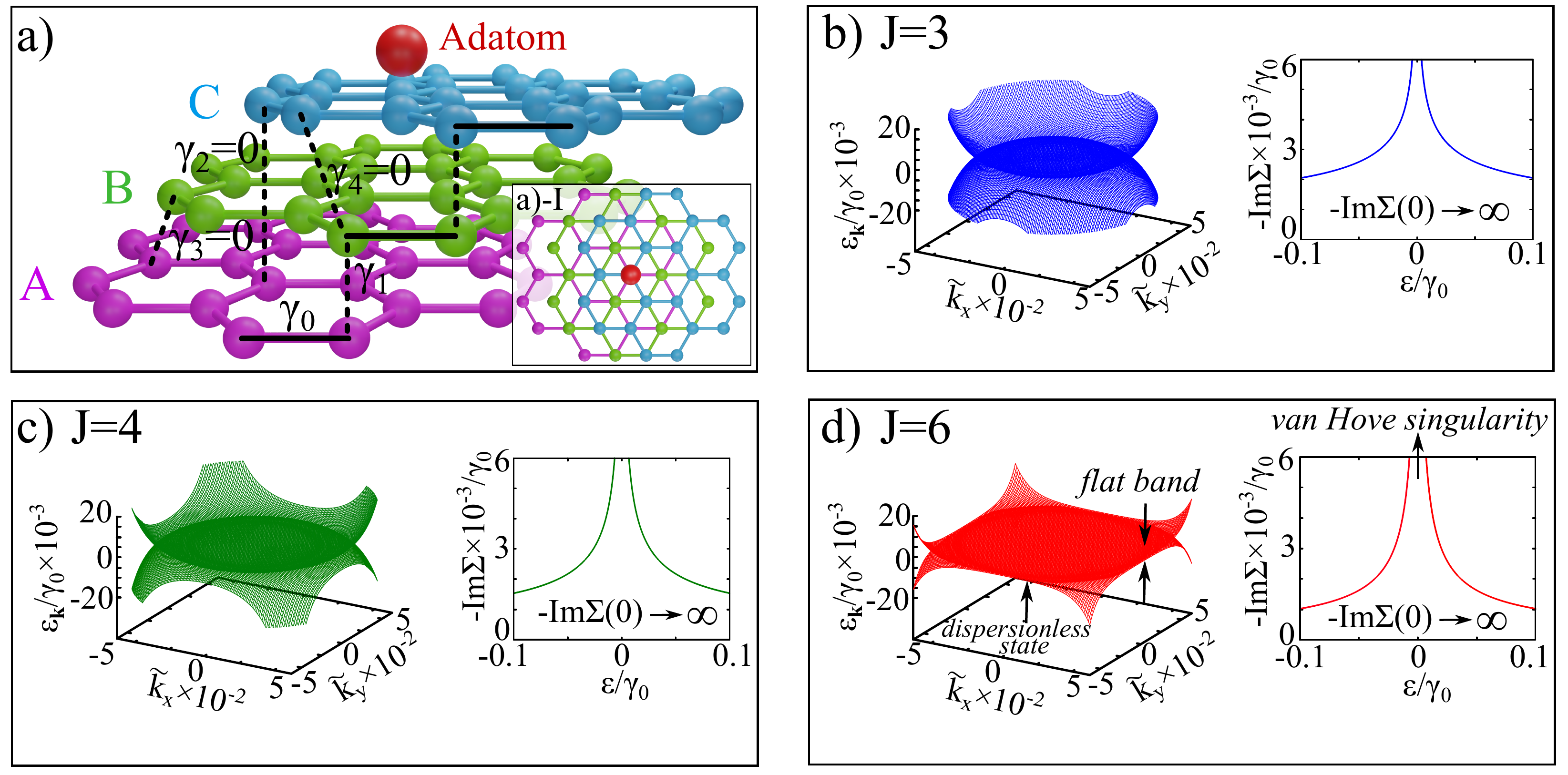}\caption{\label{fig:Pic1} (Color online) (a) Lateral view of multi-graphene
following the ABC-type stacking and an adatom on its top {[}panel
a)-I{]}. (b)-(d) Several cases $J=3,4$ and $6$, where the band-structure
{[}Eq.(\ref{eq:Dispersion}){]} and quasiparticle broadening {[}Eq.\ref{eq:ImG0}{]}
versus energy show a dispersionless state and a van Hove singularity
at Fermi energy, respectively. Upon increasing the Berry phase $J\pi$,
the flat band extension in reciprocal space enlarges too. Besides,
$J$ represents the total of monolayers\cite{MacDonald}.}
\end{figure}

Alternatively to graphene monolayer, which was already proved reliable
for mimicking the atomic collapse, we are free to ask if contemporaneous
van der Waals heterostructures could be also candidates for such,
once they are characterized by flat bands\cite{Mott-TBG,superconductivity-TBG},
i.e., the so-called van Hove singularities. However, for the entire
accomplishment of these characteristics a fine tuning of the ``twist
magic angle'' and Moiré potential conditions are demanded. Twisted
bilayer, double-bilayer and trilayer graphene\cite{Mott-TBG,superconductivity-TBG,Shen,PMID,TBGnew1,TBGnew2,TBGnew3,TBGnew4,TBGnew5,TBGnew6,TBGnew8,TDBGnew1,TDBGnew2,TDBGnew3,TDBGnew4,TDBGnew5,TDBGnew6,TBGnew7},
ABC trilayer graphene on aligned hexagonal boron nitride\cite{Calder,Chen_Signatures,Chen2019}
and transition-metal dichalcogenides\cite{Wang2020CorrelatedEP,TMD1,TMD2,TMD3,TMD4,TMD5}
compose a set of the most common examples in this research field.

In order to overcome such a technical challenge, one way is to take
into account the multilayer rhombohedral (ABC) graphene\cite{MacDonald,PhysRevB.82.035409},
which naturally shows a low energy flat band without its birth in
the Moiré superlattice induced by a twist. Interestingly enough, this
is attributed to its divergent density of states (DOS) at the Fermi
level\cite{MacDonald}, where a dispersionless state resides. Thus,
the system is known as Moiréless-type.

Consequently, based on this useful experimental aspect we benefit
from it within a theoretical perspective by proposing our adatomic
collapse effect. Here, we find that the multi-graphene of ABC-type
stacking\cite{MacDonald,PhysRevB.82.035409} turns out into a novel
platform for realizing the analogous effect of the atomic collapse.
Distinctly from earlier results\cite{PMID_Wang,Mao2016,Jiang2017},
the behavior here reported is the net effect of the interplay between
Coulomb correlations and the van Hove singularity at the Fermi energy.
Interestingly enough, this singularity enters into the power law of
the host DOS, which follows $|\varepsilon-\varepsilon_{F}|^{2/J-1},$
where the Fermi energy is $\varepsilon_{F}$ and leads to a divergence
for $J\geq3,$ with the Berry phase $J\pi$.

Physically speaking, an Anderson-like adatom\cite{Anderson} on top
of the ABC multi-graphene system, as depicted in Fig.\ref{fig:Pic1}(a),
fulfills entirely the previous constraints. Simultaneously, such an
interaction imposes a gap in the adatom DOS, where we find virtual
bound states. One of them stays below the Fermi energy due to the
correlation and characterizes a trapped electron inside the valence
band, while the other corresponds to its particle-hole symmetric (PHS)
within the conduction band, thus emulating a positron-like state.
For the quantification of this nearby van Hove singularity effect,
as we shall see below, the Kramers-Kronig transformation of the quasiparticle
broadening is adopted, from where the emergence of virtual bound states
becomes clearly evident at the host.

\section{The Model}

The low-energy Hamiltonian mimicking our system{[}Fig.\ref{fig:Pic1}(a){]}
reads $\mathcal{H}=\mathcal{H}_{\text{{ch.-ABC}}}+\mathcal{H}_{\text{{Ada.}}}+\mathcal{H}_{\text{{Hyb.}}},$
where
\begin{equation}
\mathcal{H}_{\text{{ch.-ABC}}}=\sum_{\textbf{k}s}\psi_{\textbf{k}s}^{\dagger}s[D_{J}(\tilde{k}_{-}^{J}\sigma_{+}+\tilde{k}_{+}^{J}\sigma_{-})]\psi_{\textbf{k}s}\label{eq:ABCm-graph}
\end{equation}
describes the chiral (ch.) ABC multi-graphene by a set of $J\geq3$
independent monolayers as \textit{pseudospin} doublets\cite{MacDonald},
where $J\pi$ is the Berry phase, $\psi_{\textbf{k}s}^{\dagger}=(c_{\textbf{k}s\uparrow}^{\dagger}c_{\textbf{k}s\downarrow}^{\dagger})$
represents the spinor, $c_{\textbf{k}s\sigma}^{\dagger}$ ($c_{\textbf{k}s\sigma}$)
stands for the creation (annihilation) of an electron carrying momentum
$\textbf{k}$, spin $\sigma=\uparrow,\downarrow$ and valley index
$s=\pm1$ in a single\textit{ pseudospin} doublet\cite{MacDonald}.
The part $\tilde{k}_{\pm}=(k_{x}\pm ik_{y})/k_{D_{J}}$ is of nonrelativistic
type, $k_{D_{J}}=D_{J}/v_{F}$ ($D_{J\geq3}\thickapprox\gamma_{1}=0.1\gamma_{0},$
with $\gamma_{0}=3\text{{eV}}${[}Fig.\ref{fig:Pic1}(a),\cite{MacDonald})
gives the \textit{Debye-like} momentum (energy) cutoff in terms of
the Slonczewski-Weiss-McClure parametrization\cite{PhysRevB.82.035409,TriWarping},
with $v_{F}$ the Fermi velocity and $\sigma_{\pm}=\frac{1}{2}(\sigma_{x}\pm i\sigma_{y})$
the Pauli matrices. The band-structure of Eq.(\ref{eq:ABCm-graph})
is given straightforwardly by the dispersion relation
\begin{equation}
\varepsilon_{\textbf{k}s}^{\pm}=\pm sv_{F}|\tilde{k}_{+}|^{J}k_{D_{J}},\label{eq:Dispersion}
\end{equation}
wherein the frontal sign $+(-)$ corresponds to the conduction (valence)
band. Further, it is worth citing that for $J=1$ we recover one single
graphene layer\cite{Uchoa} with well-defined Dirac cones in momentum
space with $D_{J=1}\approx\gamma_{0},$ while for $J\geq3$ we have
the chiral multi-graphene, which is characterized by a highly flat
band with a van Hove singularity due to a dispersionless state at
Fermi energy.\textcolor{red}{{} }

Additionally, a single Anderson-adatom\cite{Anderson} can be described
by the Hamiltonian
\begin{equation}
\mathcal{H}_{\text{{Ada.}}}=-\frac{U}{2}+\sum_{\sigma}(\varepsilon_{d\sigma}+\frac{U}{2})n_{d\sigma}+\frac{U}{2}(\sum_{\sigma}n_{d\sigma}-1)^{2},\label{eq:Imp}
\end{equation}
where the adatom levels are $\varepsilon_{d\sigma}$ and $\varepsilon_{d\sigma}+U,$
also recognized as the bottom and top edges of the Coulomb insulating
gap, respectively and present in the adatom DOS for the Coulomb blockade
regime\cite{Flensberg}. The $U$ term, actually, represents the Coulomb
repulsion between two electrons with opposite spins $(\bar{\sigma}=-\sigma),$
as well as the gap size and $n_{d\sigma}=(n_{d\sigma})^{2}=d_{\sigma}^{\dagger}d_{\sigma}$
stands for the number operator, with $d_{\sigma}^{\dagger}$ ($d_{\sigma}$)
the corresponding creation (annihilation) operator.

According to Ref.\cite{DFTCorrelation}, in the paramagnetic regime
the van Hove singularity for $J\geq6$ turns into a narrow peak slightly
asymmetric around the Fermi energy when Coulomb correlations are taken
into account by an interacting DFT approach. In such a case, despite
$\gamma_{2},\gamma_{3}$ and $\gamma_{4}$ finite{[}Fig.\ref{fig:Pic1}(a){]}
the trigonal warping effect of the Fermi surface\cite{TriWarping}
is smooth characterized by a quasi-flat state nearby the Fermi energy\cite{DFTCorrelation}.
This scenario is qualitatively captured by Eq.(\ref{eq:ABCm-graph})
as we shall see below.

The hybridization term stands for an adatom for the top site-type
geometry, i.e., right on a carbon atom such as Au and Sn prefer\cite{AdatomAbove}.
This leads to
\begin{equation}
\mathcal{H}_{\text{{Hyb.}}}=v(\frac{1}{\sqrt{\mathcal{N}}}\sum_{\textbf{k}s\sigma}c_{\textbf{k}s\sigma}^{\dagger}d_{\sigma}+\text{{H.c.}),}\label{eq:Hyb}
\end{equation}
with $v$ as the host-adatom hopping term and $\mathcal{N}$ is the
number of states delimited by the Debye radius $k_{D_{J}}.$ However,
the hollow (center of the hexagon ring) and bridge site-types (middle
of C-C bond) are other possibilities for adsorption\cite{AdatomAbove}
to be explored elsewhere.

{In our Anderson-like Hamiltonian $\mathcal{H}$\cite{Anderson}
for $J\geq3$ we introduce atomic collapse regimes similarly to Ref.\cite{Criteria}:}\textcolor{red}{{}
}(i) $v\equiv0$ is enough to define the subcritical limit without
host resonant states; (ii) charge criticality is the starting point
of $\mathcal{H}$ wherein $v\neq0$ naturally and $\varepsilon_{d\sigma}=U=0$
stands for the PHS regime without correlation. As $f_{0\sigma}\equiv\frac{1}{\sqrt{\mathcal{N}}}\sum_{\textbf{k}s}c_{\textbf{k}s\sigma}$
in Eq.(\ref{eq:Hyb}) acts as a host localized dispersionless state
at Fermi energy coupled to $d_{\sigma},$ the so-called bonding (adatomic
collapse) and antibonding (positron-like) molecular states symmetrically
below and above such an energy emerge, respectively; (iii) in the
interacting PHS regime, the super criticality emerges by burying $\varepsilon_{d\sigma}=-\frac{U}{2}\neq0$
into the Fermi sea. It enhances the effective charge of the adatom
in analogy to calcium ions gradually deposited on graphene monolayer
up to electronic localization\cite{PMID_Wang,CaIons2}. Consequently,
we observe two extra PHS localizations emerging within the Coulomb
insulating gap at energies $-|\varepsilon_{d\sigma}|<\varepsilon<0$
and $0<\varepsilon<|\varepsilon_{d\sigma}|$ as the novel adatomic
collapse and positron-like states, respectively.

{Simultaneously in (ii) and (iii) }\textit{{Friedel-like}}{{}
oscillations exhibit the same spatial dependence as the long-range
Coulomb interaction (\cite{Friedel} and out of the present scope).}
We show that the relativistic linear dispersion of single graphene
is needless and that the van Hove singularity is pivotal for the effect.

\subsection{The Adatom DOS and LDOS}

From the time Fourier transform $\tilde{\mathcal{G}}_{\text{{Ada.}}\sigma}$
of the adatom retarded Green's function (GF)\cite{Flensberg} $\mathcal{G}_{\text{{Ada.}}\sigma}=-i\theta(t)\left\langle \left\{ d_{\sigma}(t),d_{\sigma}^{\dagger}(0)\right\} \right\rangle _{\mathcal{H}},$
we define the adatom DOS
\begin{equation}
\text{{DOS}}=-\frac{1}{\pi}\text{{Im}}\tilde{\mathcal{G}}_{\text{{Ada.}}\sigma}.\label{eq:DOS}
\end{equation}
We consider the Coulomb blockade regime\cite{Flensberg}, in which
the correlation $U$ in $\tilde{\mathcal{G}}_{\text{{Ada.}}\sigma}$
is accounted for within the framework of the Hubbard-I approximation\cite{Flensberg},
which leads to the well-known adatom GF\cite{TopoFano}:
\begin{equation}
\tilde{\mathcal{G}}_{\text{{Ada.}}\sigma}=\frac{1}{(-\text{{Im}}\Sigma)}\left(\frac{w_{x}}{x+i}+\frac{w_{\bar{x}}}{\bar{x}+i}\right).\label{eq:Hubbard}
\end{equation}
This expression introduces in Eq.(\ref{eq:DOS}) the Hubbard bands
centered around the gap edges $\varepsilon_{d\sigma}$ and $\varepsilon_{d\sigma}+U,$
with quasiparticle broadening $-\text{{Im}}\Sigma=-v^{2}\text{{Im}\ensuremath{\tilde{\mathcal{G}}_{\sigma}^{0}}}$
each\cite{Flensberg}, wherein $w_{x}=1-\left\langle n_{d\bar{\sigma}}\right\rangle $
and $w_{\bar{x}}=1-w_{x}$ represent the spectral weights (peak amplitudes)
of the bands, respectively.

Additionally, $\left\langle n_{d\sigma}\right\rangle =\int_{-D}^{0}\text{{DOS}}d\varepsilon$
is the electronic occupation of the adatom and
\begin{align}
\tilde{\mathcal{G}}_{\sigma}^{0} & =\frac{1}{\mathcal{N}}\sum_{\textbf{k}s}\frac{\varepsilon+i0^{+}}{(\varepsilon+i0^{+})^{2}+(\varepsilon_{\textbf{k}s}^{+})^{2}}\nonumber \\
\label{eq:PHostGF}
\end{align}
is the pristine host GF for the ABC multi-graphene, which allows to
define the natural Fano coordinates\cite{TopoFano}
\begin{equation}
x=\frac{\varepsilon-\varepsilon_{d\sigma}-\text{Re}\Sigma}{(-\text{{Im}}\Sigma)}\label{eq:naturalx}
\end{equation}
and
\begin{equation}
\bar{x}=\frac{\varepsilon-\varepsilon_{d\sigma}-U-\text{Re}\Sigma}{(-\text{{Im}}\Sigma)},\label{eq:naturalxbar}
\end{equation}
where $\text{Re}\Sigma$ is the Kramers-Kronig transformation of $-\text{{Im}}\Sigma.$
By applying such a transformation to $-\text{{Im}}\Sigma,$ we obtain
\begin{align}
\text{{Re}\ensuremath{\Sigma}} & =\frac{1}{\pi}\int_{-D_{J}}^{+D_{J}}\frac{(-\text{{Im}}\Sigma)}{\varepsilon-y}dy\nonumber \\
 & =q_{J}(-\text{{Im}}\Sigma),\label{eq:KramersKronig}
\end{align}
wherein $y=u\varepsilon$ and $-\text{{Im}}\Sigma$ is the quasiparticle
broadening modulated by the asymmetry parameter (also called as Fano
parameter\cite{Fano})
\begin{align}
q_{J} & =\frac{1}{\pi}\text{{sgn}}(\varepsilon)\text{{P.V.}}\int_{-D_{J}/\varepsilon}^{+D_{J}/\varepsilon}\frac{|u|^{2/J-1}}{1-u}du,\label{eq:FanoqGeral}
\end{align}
where \text{P.V.} stands for the Cauchy principal value. This object,
as we will see later on, is profoundly connected to the emergence
of the virtual bound states at the host.

After applying to {Eq.(\ref{eq:PHostGF})} the following methodology:
(i) the standard substitution $\mathcal{N}=\sum_{\textbf{k}s}\rightarrow\frac{\mathcal{A}}{(2\pi)^{2}}\int d^{2}\textbf{k}=\frac{\mathcal{A}}{2\pi}k_{D_{J}}^{2}$,
with $\mathcal{A}$ as the area element in real space; (ii) the hyper-polar
transformation given by $k_{x}=k_{D_{J}}\left(\frac{\varepsilon_{\textbf{k}s}^{+}}{D_{J}}\right)^{\frac{1}{J}}\cos\theta$
and $k_{y}=k_{D_{J}}\left(\frac{\varepsilon_{\textbf{k}s}^{+}}{D_{J}}\right)^{\frac{1}{J}}\sin\theta$
($0\leq\theta\leq2\pi$), with Jacobian $\mathcal{J}(\varepsilon_{\textbf{k}s}^{+},\theta)=\frac{k_{D_{J}}^{2}}{J}\frac{(\varepsilon_{\textbf{k}s}^{+})^{2/J-1}}{D_{J}^{2/J}}$
and property $\int\tilde{\mathcal{G}}_{\sigma}^{0}d^{2}\textbf{k}=\int\tilde{\mathcal{G}}_{\sigma}^{0}\mathcal{J}(\varepsilon_{\textbf{k}s}^{+},\theta)d\varepsilon_{\textbf{k}s}^{+}d\theta,$
we can determine the quasiparticle broadening as follows

\begin{equation}
-\text{{Im}}\Sigma=v^{2}\frac{\pi\left|\varepsilon\right|^{2/J-1}}{JD_{J}^{2/J}},\label{eq:ImG0}
\end{equation}
with scaling-law $\left|\varepsilon\right|^{2/J-1}$ exhibiting a
van Hove singularity at Fermi level $\varepsilon=0$ for $J\geq3.$
We do not discuss $J=2$, since it corresponds to metals without divergence.

We remark that for the situation $q_{J=1}$ (single graphene), the
function in Eq.(\ref{eq:FanoqGeral}) that depends on $u$ does not
vanish in the boundaries $u\rightarrow\pm\infty.$ This feature imposes
a mathematical challenge for the integral calculation and to handle
it accordingly, we just follow Ref.\cite{Uchoa} for single graphene,
but with the cutoff $D_{J=1}\approx\gamma_{0}$ instead. We solve
first the integral analytically by keeping $D_{J=1}/\varepsilon$
finite and the limit $\varepsilon/D_{J=1}\ll1$ is only assumed in
the numerical evaluations of $\text{{Re}\ensuremath{\Sigma}}$. Consequently,
\begin{equation}
\text{{Re}\ensuremath{\Sigma}}(J=1)=v^{2}\frac{\varepsilon}{D_{J=1}^{2}}\ln\frac{\left|\varepsilon^{2}\right|}{\left|\varepsilon^{2}-D_{J=1}^{2}\right|}\label{eq:J1}
\end{equation}
emerges as dependent on energy for $\varepsilon/D_{J=1}\ll1.$ Nevertheless,
the pursued vanishing characteristic in $q_{J}$ as a function of
$u$ exists for $q_{J\geq3}$ and as aftermath, the integral in Eq.(\ref{eq:FanoqGeral})
does behave finite for $D_{J\geq3}/\varepsilon\rightarrow\infty.$
This yields a capital result, i.e., an asymmetry parameter discretized
in the number of the system layers $J$, which reads
\begin{equation}
q_{J\geq3}=\text{{sgn}}(\varepsilon)\cot(C_{2J}),\label{eq:FanoqMaior3}
\end{equation}
where we define $C_{2J}\equiv(360^{\circ}/2J)$ as the angle for the
corresponding rotational symmetry group.

Additionally, in order to understand also the emergence of virtual
bound states in the host LDOS coupled to the adatom and its interplay
with the van Hove singularity, we should consider the following Green's
function
\begin{equation}
\mathcal{G}_{\sigma}=-i\theta(t)\left\langle \left\{ \frac{1}{\sqrt{\mathcal{N}}}\sum_{\textbf{k}s}c_{\textbf{k}s\sigma}(t),\frac{1}{\sqrt{\mathcal{N}}}\sum_{\textbf{k}s}c_{\textbf{k}s\sigma}^{\dagger}(0)\right\} \right\rangle _{\mathcal{H}},\label{eq:GFfo}
\end{equation}
which yields
\begin{align}
\text{{LDOS}} & =-\frac{1}{\pi}\text{{Im}}\tilde{\mathcal{G}}_{\sigma}=\frac{(-\text{{Im}}\Sigma)}{\pi v^{2}}\sum_{p=x,\bar{x}}w_{p}\frac{(p+q_{J})^{2}}{p^{2}+1}\label{eq:LDOS}
\end{align}
in terms of the Fano formula\cite{Fano}, where $\tilde{\mathcal{G}}_{\sigma}$
is the time Fourier transform of $\mathcal{G}_{\sigma}.$ To conclude,
if we subtract the background $(-\text{{Im}}\Sigma)/\pi v^{2}$ from
Eq.(\ref{eq:LDOS}), i.e., we obtain the adatom induced LDOS, which
reads
\begin{equation}
\Delta\text{{LDOS}}=\text{{LDOS}}-\frac{(-\text{{Im}}\Sigma)}{\pi v^{2}}.\label{eq:DeltaLDOS}
\end{equation}

\section{Results and discussion}

Here we present the counterpart of the atomic collapse in relativistic
atomic Physics, i.e., the adatomic collapse introduced by us. For
a sake of simplicity, we take into account the model PHS regime $\varepsilon_{d\sigma}=\varepsilon_{d}=-\frac{U}{2}=-0.07\gamma_{0}$
and $w_{x}=w_{\bar{x}}=1/2$ for the paramagnetic solution $\left\langle n_{d\sigma}\right\rangle =1/2$
due to a self-consistent calculation and $v/D_{J}=0.14$.

\begin{figure}[!]
\centering\includegraphics[width=1\columnwidth]{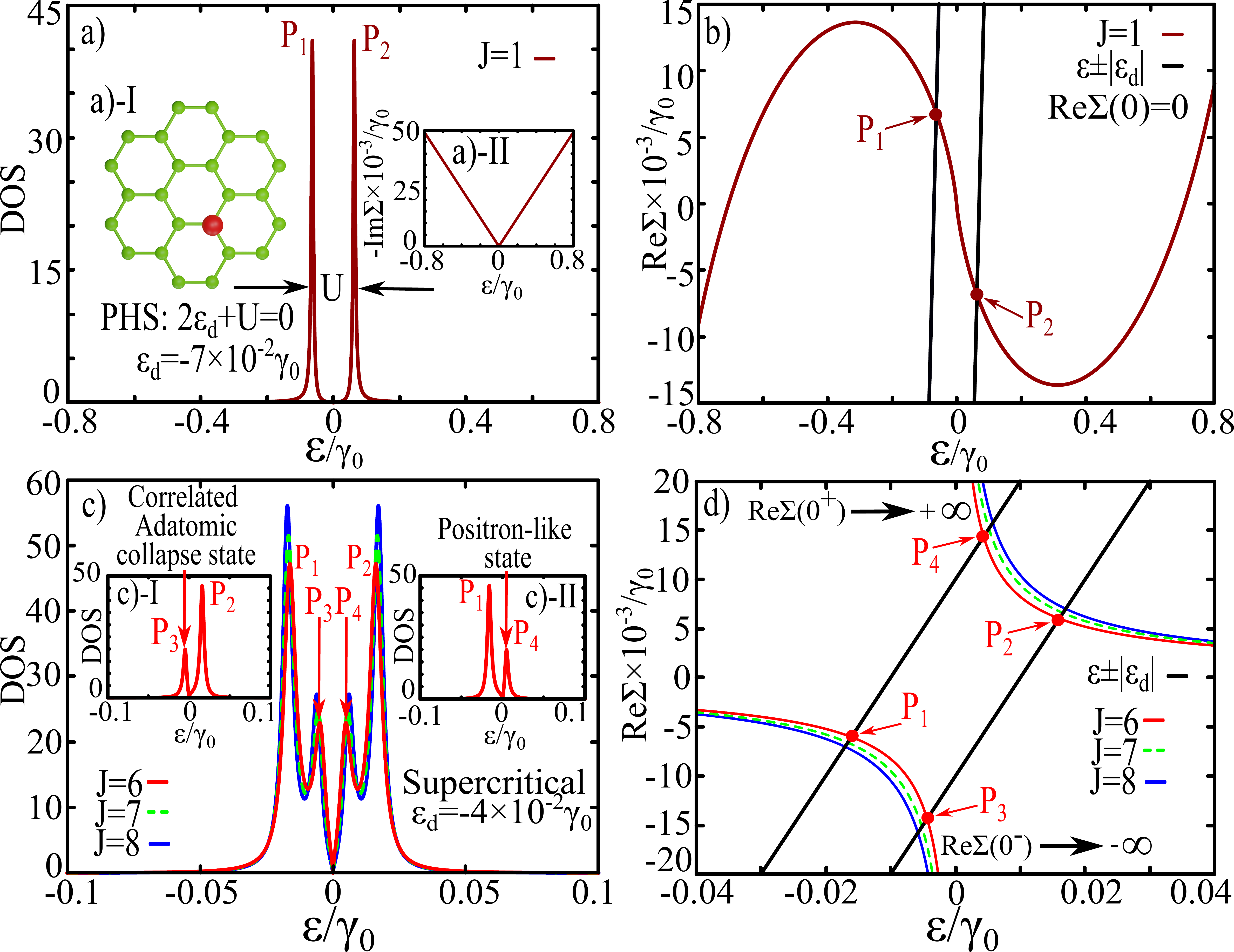}\caption{\label{fig:Pic2} (Color online) Particle-hole symmetry (PHS) is taken
into account. (a) Single graphene with an adatom {[}a)-I{]}, linear
quasiparticle broadening on energy with a pseudogap at Fermi level
{[}a)-II{]} and typical Coulomb gap in the adatom DOS. (b) Gap edges
appear as P1 and P2 in the Kramers-Kronig transformation of $-\text{{Im}}\Sigma,$
namely $\text{{Re}}\Sigma.$ (c) Adatom DOS for $J=6,7$ and 8 exhibiting
inner virtual bound states to the gap marked by P3 and P4 solely for
$J=6.$ (d) As for the ABC stacking $\text{{Re}}\Sigma$ diverges
at Fermi energy, the bottom P1 and top P2 Coulomb gap edges lead to
virtual bound states P4 and P3, respectively. The resonant peak P3
is due to the correlation $U$ and mimics the atomic collapse, once
an electron state localizes at $\varepsilon<0.$ In contrast, P4 is
the particle-hole symmetric of P3 at $\varepsilon>0,$ analogously
to a positron state.}
\end{figure}

By considering Eq.(\ref{eq:DOS}) in Figs.\ref{fig:Pic2}(a) and (c)
we contrast the spectral analysis of the adatom DOS for single graphene
$(J=1)$ and several ABC multi-graphene systems $(J\geq6)$, respectively.
In Fig.\ref{fig:Pic2}(a) we clearly notice the Hubbard bands around
$\varepsilon\approx-|\varepsilon_{d}|$ (P1 or bottom edge of the
gap) and $\varepsilon\approx+|\varepsilon_{d}|$ (P2 or top edge)
together with a pseudogap at Fermi energy in the quasiparticle broadening,
as depicted in the inset a)-II. By imposing $x=0$ and $\bar{x}=0$
in Eqs.(\ref{eq:naturalx}) and (\ref{eq:naturalxbar}) we obtain
$\varepsilon+|\varepsilon_{d}|=\text{{Re}\ensuremath{\Sigma}}$ and
$\varepsilon-|\varepsilon_{d}|=\text{{Re}\ensuremath{\Sigma},}$ respectively.

As a result, we are able to determine the poles of Eq.(\ref{eq:Hubbard})
thus providing the precise positions of the insulating gap edges.
The latter appear graphically in Fig.\ref{fig:Pic2}(b), where we
see two crossing points between the black and wine{[}Eq.(\ref{eq:J1}){]}
curves, once $\text{{Re}\ensuremath{\Sigma}}$ just has van Hove singularities
far away the Fermi energy and located at energy cutoffs $\pm\gamma_{0}${[}Eq.(\ref{eq:J1}){]}.
In such a domain interceptions do not occur. Additionally, $\text{{Re}\ensuremath{\Sigma(0)}}$
shows a pseudogap too and, most importantly, $\text{{Re}\ensuremath{\Sigma}}$
follows the same energy sign dependence of $-\text{{Im}}\Sigma$.

The scenario changes drastically by considering the ABC multi-graphene
with $J\geq6$ as depicted in Fig.\ref{fig:Pic2}(c) due to the asymmetry
parameter $q_{J\geq6}$ of Eq(\ref{eq:FanoqMaior3}). In this analysis
we verify the expected points P1 and P2, i.e., the Coulomb gap edges.
Counterintuitively, two inner resonant states to the gap represented
by P3 and P4 appear. Such points describe emerging virtual bound states
arising from the Moiréless characteristic of the ABC multi-graphene
system.

The aforementioned feature is due to the dispersionless state pinned
at the Fermi level according to the band-structures in inset panels
(b)-(d) of Fig.\ref{fig:Pic1}, which can be obtained from Eq.(\ref{eq:Dispersion}).
In Fig.\ref{fig:Pic1} we relax the condition $J\geq6$ just for a
sake of clearness. It means that such a flat band manifests as a van
Hove divergence in the quasiparticle broadening exactly at Fermi energy{[}Eq.(\ref{eq:ImG0})
and inset panels (b)-(d) of Fig.\ref{fig:Pic1}{]}, as well as in
its Kramers-Kronig transformation{[}Eq.(\ref{eq:KramersKronig}){]}.
However, the way that such a singularity presents itself in the latter
quantity is quite curious. Differently from $-\text{{Im}\ensuremath{\Sigma}}(0)\rightarrow\infty$
for the quasiparticle broadening{[}Eq.(\ref{eq:ImG0}){]} we have
$\text{{Re}\ensuremath{\Sigma}}(0^{-})\rightarrow-\infty$ and $\text{{Re}\ensuremath{\Sigma}}(0^{+})\rightarrow+\infty$
instead, exclusively because there is the energy asymmetry factor
$\text{{sgn}}(\varepsilon)$ in the parameter $q_{J\geq6}$ of Eq.(\ref{eq:FanoqMaior3}).
It reverses the energy sign of $-\text{{Im}\ensuremath{\Sigma}}$
when we swap the valence by the conduction bands. This can be visualized
in Fig.\ref{fig:Pic2}(d), where two extra crossing points are recorded,
i.e., P3 and P4. Noteworthy, we realize that such new poles are correlated
to the Coulomb gap edges: the black line above the van Hove singularity
contains P1(bottom edge) and P4, while the corresponding below has
P2(top edge) and P3. Equivalently, the existence of the virtual bound
state P3 is correlated to the Coulomb gap edge P2.

As the peak P3 describes an electronic localization beneath the Fermi
level, later on, we will demonstrate that its formation emulates the
counterpart of the atomic collapse phenomenon found in relativistic
atomic Physics. Particularly, P3 arises from the correlation $U,$
once such a resonant state is derived from the solution of $\bar{x}=0${[}Eq.(\ref{eq:naturalxbar}){]}
and gives the top edge of the gap. To perceive this characteristic
at the inset panel c)-I of Fig.\ref{fig:Pic2} the corresponding adatom
DOS part of Eq.(\ref{eq:Hubbard}) with spectral weight $w_{\bar{x}}$
is presented. In such an inset, we can see explicitly the electronic
localization below the Fermi energy induced by the correlation $U.$

\begin{figure}[!]
\centering\includegraphics[width=1\columnwidth]{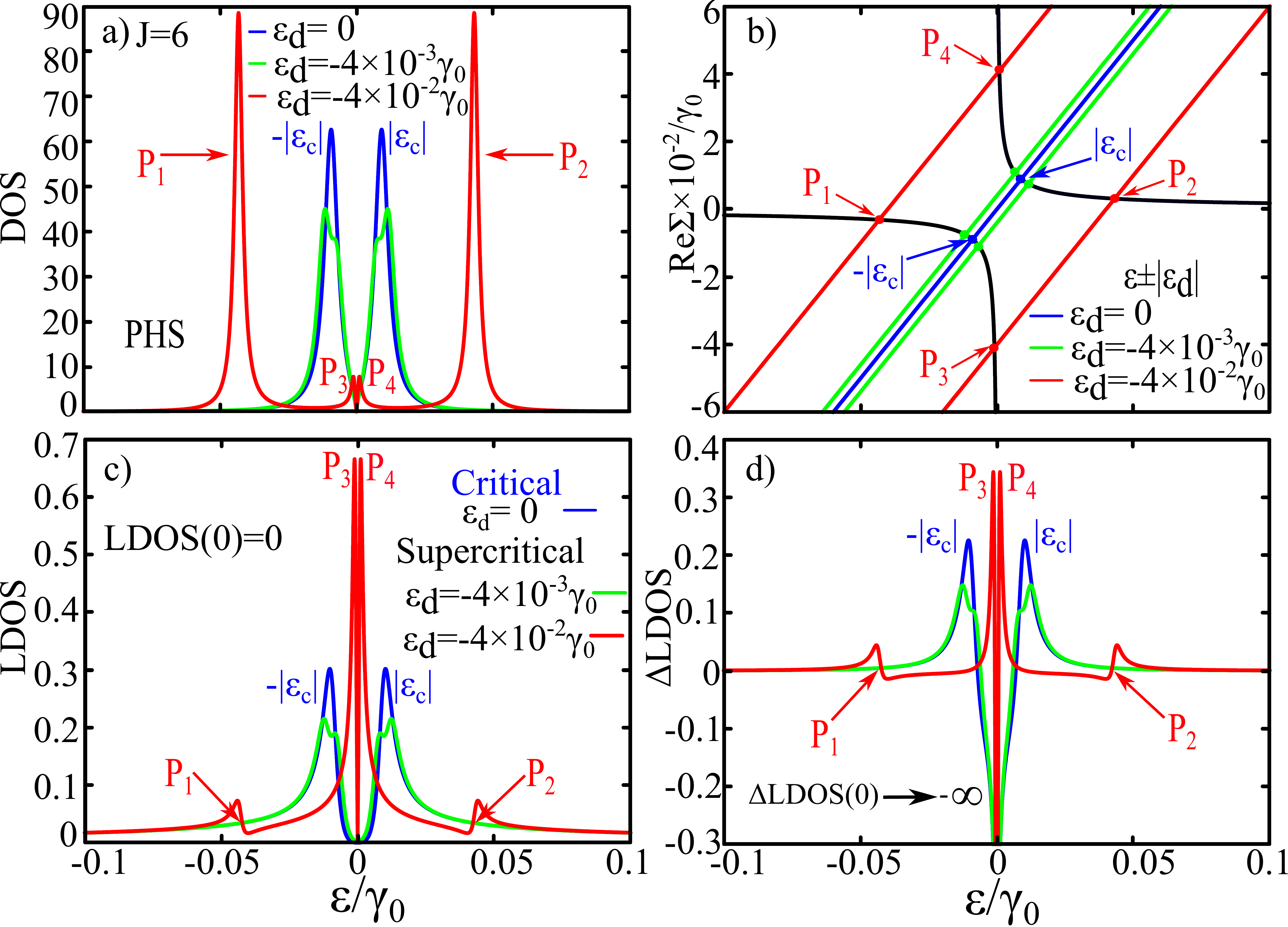}\caption{\label{fig:Pic3} {(Color online) Mechanism of the adatomic collapse
in the ABC stacking: (a) Adatom DOS in the PHS regime for several
energies $\varepsilon_{d},$ where quenching of virtual bond states
occurs upon burying the adatom energy within the Fermi sea of the
ABC multi-graphene. (b) $\text{{Re}}\Sigma$ reveals that more deeper
is $\varepsilon_{d},$ more the states P3 and P4 fall into the van
Hove singularity and couple to the host. (c) These states appear in
the system LDOS, where the singularity is absent. (d) This lacking
is due to the induced $\Delta\text{{LDOS}}$ wherein a negative van
Hove singularity cancels the one positive in $(-\text{{Im}}\Sigma)/\pi v^{2}.$}
The critical energy is $|\varepsilon_{c}|\sim9\times10^{-3}\gamma_{0},$
where $-(+)$ for the adatomic collapse (positron-like) state and{{}
atomic number $\mathcal{Z}_{c}\sim0.96.$}}
\end{figure}

Additionally, the state P4 is the particle-hole symmetric of P3 obtained
from the condition $x=0$ in Eq.(\ref{eq:naturalx}) and plays the
role of a positron state, since its localization occurs above the
Fermi energy. This feature can be found at the inset panel c)-II of
Fig.\ref{fig:Pic2}, but with the corresponding DOS part carrying
the spectral weight $w_{x}.$ With this in mind, in Fig.\ref{fig:Pic3}
we reveal the underlying mechanism responsible for the aforementioned
effect in the ABC multi-graphene.

The critical limit is depicted by the blue curves in Figs.\ref{fig:Pic3}(a)-(d)
with $\varepsilon_{d}=U=0$ and energies $\varepsilon=\mp|\varepsilon_{c}|=\mp[\sqrt{C_{2J}q_{J}(\varepsilon>0)}(\frac{v}{D_{J}})]^{\frac{J}{J-1}}D_{J}$
obtained from $x=\bar{x}=0${[}Eqs.(\ref{eq:naturalx}) and (\ref{eq:naturalxbar}){]},
where $-(+)$ stands for the adatomic collapse (positron-like) state.
Particularly for $J=6$ we find $|\varepsilon_{c}|\sim9\times10^{-3}\gamma_{0}=2.7\times10^{-2}\text{{eV}}.${{}
Here we choose to mimic the adatom as a highly excited effective hydrogen-like
atom, which naturally imposes the upper critical atomic number $\mathcal{Z}_{c}^{\text{{upper}}}=1.$
In this way, we point out that $n=22$ provides the very close value
$|\varepsilon_{c}^{\text{{upper}}}|=\mathcal{Z}_{c}^{\text{{upper}}}\frac{13.6\text{{eV}}}{n^{2}}\sim2.8\times10^{-2}\text{{eV}}$
for the critical energy $|\varepsilon_{c}|.$ If we maintain $|\varepsilon_{c}^{\text{{upper}}}|$
in such a formula, the breakdown of this physics picture is expected
to occur for bigger values of $n,$ i.e., $\mathcal{Z}_{c}^{\text{{upper}}}\gg1.$
Thus, to match accurately with $|\varepsilon_{c}|$ and then find
an effective critical atomic number $\mathcal{Z}_{c}<1,$ we keep
$n$ instead and perform the evaluation $\mathcal{Z}_{c}=\frac{n^{2}|\varepsilon_{c}|}{13.6\text{{eV}}}\sim0.96,$
which is certainly not for an ultra-heavy nucleus. Similarly, in Ref.\cite{PMID_Wang}
the value is $\mathcal{Z}_{c}\sim0.25.$ }

Fig.\ref{fig:Pic3}(a) shows the analysis of the adatom DOS for several
values of $\varepsilon_{d}=-\frac{U}{2}$ with $J=6,$ where we can
clearly perceive that by burying the adatom level into the Fermi sea
of the host, the P3 and P4 peaks become suppressed with their positions
shifted towards the Fermi energy, where the van Hove singularity resides.
In the other hand, the Coulomb gap edges P1 and P2 exhibit enhanced
amplitudes with peak positions displaced far away the singularity.
Thus, in order to understand the aforementioned novel behavior, we
present the adatom energy dependence of $\text{{Re}\ensuremath{\Sigma}}$
in Fig.\ref{fig:Pic3}(b). Interestingly enough, this panel reveals
that the poles P3 and P4 dive into the van Hove singularity upon decreasing
$\varepsilon_{d}.$

Fig.\ref{fig:Pic3}(b) elucidates that there is a sharp region in
energy domain sufficiently close to the Fermi level that bonds an
electron as a state in the host energy continuum {[}panel (c){]},
as well as traps an electron within the adatom site as a virtual bond
state with energy below the Fermi level. Additionally, such a trapping
leads to a resonant state above the Fermi energy, which mimics a positron
state in analogy to its counterpart in relativistic atomic Physics.

When such a situation is established, we realize the adatomic collapse
introduced by us. It means that when such poles are finally captured
by the singularity, these states inevitably localize in the host band
{[}see P3, P4 in the LDOS given by Eq.(\ref{eq:LDOS}) and depicted
in panel (c){]}. Surprisingly, as we have the negative van Hove singularity
in $\Delta\text{{LDOS}}(0)\rightarrow-\infty$ shown in Fig.\ref{fig:Pic3}(d),
which interferes destructively with the corresponding positive in
$-\text{{Im}\ensuremath{\Sigma}}(0)/\pi v^{2}\rightarrow\infty,$
we find $\text{{LDOS}}(0)=0$ in Fig.\ref{fig:Pic3}(c).

As expected, if we extrapolate the analysis away from the PHS regime
$\varepsilon_{d}\neq-\frac{U}{2}$ in Figs.\ref{fig:Pic2} and \ref{fig:Pic3}
the states P3 and P4 would exhibit distinct spectral amplitudes without
mirror symmetry around the Dirac point. Such a feature is derived
straightforwardly from the molecular binding in Eq.(\ref{eq:Hyb})
with the dispersionless states $\varepsilon_{d}$ and $\varepsilon_{d}+U$
for the orbital $d_{\sigma}$ flanking asymmetrically the observed
in $f_{0\sigma}\equiv\frac{1}{\sqrt{\mathcal{N}}}\sum_{\textbf{k}s}c_{\textbf{k}s\sigma}$
at the Dirac point. This mimics the Fermi energy off the Dirac point
situation and independently, the adatomic collapse would persist.
In summary, according to Fig.\ref{fig:Pic3}(b) in this generalized
situation P3 and P4 would emerge from the crossings $\varepsilon-(U-|\varepsilon_{d}|)=\text{{Re}\ensuremath{\Sigma}}$
and $\varepsilon+|\varepsilon_{d}|=\text{{Re}\ensuremath{\Sigma}},$
respectively.

\section{Conclusions }

In this work we find virtual bound states within the Coulomb insulating
gap of an adatom coupled to a multi-graphene system obeying the ABC
stacking. We demonstrate that one of the virtual states mimics the
atomic collapse phenomenon known in relativistic atomic Physics\cite{SHYTOV2009,JR1977}.
Interestingly enough, the second virtual state reported corresponds
to the particle-hole symmetric of the latter, thus playing the role
of the positron state. In this way, we introduce our adatomic collapse,
which arises from the interplay between the Moiréless feature of the
multi-graphene host and the Coulomb correlations within the Anderson-like
adatom. {In this manner we estimate the effective
critical atomic number $\mathcal{Z}_{c}\sim0.96$\cite{PMID_Wang}.}
Therefore, our findings make explicit that the ABC multi-graphene
system is a rich environment for strongly correlated phenomena and
emulation of relativistic Physics.

\section{Acknowledgments }

We thank the Brazilian funding agencies CNPq (Grants Nr. 302887/2020-2
and 308695/2021-6), Coordenação de Aperfeiçoamento de Pessoal de Nível
Superior - Brasil (CAPES) -- Finance Code 001 and the São Paulo Research
Foundation (FAPESP Grant No. 2018/09413-0). We thank Prof. Edson Vernek
for discussions of the adatomic collapse in terms of a molecular binding,
which helped us to better explain the effect.

\bibliographystyle{unsrt}

\begin{thebibliography}{10}

\bibitem{SHYTOV2009}
Andrei Shytov, Mark Rudner, Nan Gu, Mikhail Katsnelson, and Leonid Levitov.
\newblock Atomic collapse, lorentz boosts, klein scattering, and other
  quantum-relativistic phenomena in graphene.
\newblock {\em Solid State Communications}, 149(27):1087--1093, 2009.
\newblock Recent Progress in Graphene Studies.

\bibitem{JR1977}
J~Reinhardt and W~Greiner.
\newblock Quantum electrodynamics of strong fields.
\newblock {\em Reports on Progress in Physics}, 40(3):219, mar 1977.

\bibitem{Schweppe1983}
J.~Schweppe, A.~Gruppe, K.~Bethge, H.~Bokemeyer, T.~Cowan, H.~Folger, J.~S.
  Greenberg, H.~Grein, S.~Ito, R.~Schule, D.~Schwalm, K.~E. Stiebing,
  N.~Trautmann, P.~Vincent, and M.~Waldschmidt.
\newblock Observation of a peak structure in positron spectra from u+cm
  collisions.
\newblock {\em Phys. Rev. Lett.}, 51:2261--2264, Dec 1983.

\bibitem{Cowan1985}
T.~Cowan, H.~Backe, M.~Begemann, K.~Bethge, H.~Bokemeyer, H.~Folger, J.~S.
  Greenberg, H.~Grein, A.~Gruppe, Y.~Kido, M.~Kl\"uver, D.~Schwalm,
  J.~Schweppe, K.~E. Stiebing, N.~Trautmann, and P.~Vincent.
\newblock Anomalous positron peaks from supercritical collision systems.
\newblock {\em Phys. Rev. Lett.}, 54:1761--1764, Apr 1985.

\bibitem{PMID_Wang}
Yang Wang, Dillon Wong, and Andrey V. et~al. Shytov.
\newblock Observing atomic collapse resonances in artificial nuclei on
  graphene.
\newblock {\em Science (New York, N.Y.)}, 340(6133):734--737, May 2013.

\bibitem{CaIons2}
Alireza Saffarzadeh and George Kirczenow.
\newblock Coulomb bound states and resonances due to groups of ca dimers
  adsorbed on suspended graphene.
\newblock {\em Phys. Rev. B}, 90:155404, Oct 2014.

\bibitem{Mao2016}
Jinhai Mao, Yuhang Jiang, and Dean et~al. Moldovan.
\newblock Realization of a tunable artificial atom at a supercritically charged
  vacancy in graphene.
\newblock {\em Nature Physics}, 12:545--549, 2016.

\bibitem{Jiang2017}
Yuhang Jiang, Jinhai Mao, and Dean et~al. Moldovan.
\newblock Tuning a circular p-n junction in graphene from quantum confinement
  to optical guiding.
\newblock {\em Nature Nanotechnology}, 12:1045--1049, 2017.

\bibitem{MacDonald}
Hongki Min and Allan~H. MacDonald.
\newblock Electronic structure of multilayer graphene.
\newblock {\em Progress of Theoretical Physics Supplement}, 176:227--252, 06
  2008.

\bibitem{Mott-TBG}
Yuan Cao, Valla Fatemi, Ahmet Demir, Shiang Fang, Spencer~L. Tomarken, Jason~Y.
  Luo, Javier~D. Sanchez-Yamagishi, Kenji Watanabe, Takashi Taniguchi,
  Efthimios Kaxiras, Ray~C. Ashoori, and Pablo Jarillo-Herrero.
\newblock Correlated insulator behaviour at half-filling in magic-angle
  graphene superlattices.
\newblock {\em Nature}, 556:7699, 2018.

\bibitem{superconductivity-TBG}
Y.~Cao, S.~Fatemi, V.and~Fang, Kenji Watanabe, Takashi Taniguchi, Efthimios
  Kaxiras, and Pablo Jarillo-Herrero.
\newblock Unconventional superconductivity in magic-angle graphene
  superlattices.
\newblock {\em Nature}, 556:7699, Feb 2018.

\bibitem{Shen}
Cheng Shen, Yanbang Chu, QuanSheng Wu, Na~Li, Shuopei Wang, Yanchong Zhao, Jian
  Tang, Jieying Liu, Jinpeng Tian, Kenji Watanabe, Takashi Taniguchi, Rong
  Yang, Zi~Yang Meng, Dongxia Shi, Oleg~V. Yazyev, and Guangyu Zhang.
\newblock Correlated states in twisted double bilayer graphene.
\newblock {\em Nature Physics}, 16:520--525, 2020.

\bibitem{PMID}
Yuan Cao, Daniel Rodan-Legrain, Oriol Rubies-Bigorda, Jeong~Min Park, Kenji
  Watanabe, Takashi Taniguchi, and Pabl Jarillo-Herrero.
\newblock Tunable correlated states and spin-polarized phases in twisted
  bilayer-bilayer graphene.
\newblock {\em Nature}, 583:215--220, 2020.

\bibitem{TBGnew1}
Yuan Cao, Debanjan Chowdhury, Daniel Rodan-Legrain, Oriol Rubies-Bigorda, Kenji
  Watanabe, Takashi Taniguchi, T.~Senthil, and Pablo Jarillo-Herrero.
\newblock Strange metal in magic-angle graphene with near planckian
  dissipation.
\newblock {\em Phys. Rev. Lett.}, 124:076801, Feb 2020.

\bibitem{TBGnew2}
Ming Xie and A.~H. MacDonald.
\newblock Weak-field hall resistivity and spin-valley flavor symmetry breaking
  in magic-angle twisted bilayer graphene.
\newblock {\em Phys. Rev. Lett.}, 127:196401, Nov 2021.

\bibitem{TBGnew3}
Grigory Tarnopolsky, Alex~Jura Kruchkov, and Ashvin Vishwanath.
\newblock Origin of magic angles in twisted bilayer graphene.
\newblock {\em Phys. Rev. Lett.}, 122:106405, Mar 2019.

\bibitem{TBGnew4}
Moon~Jip Park, Youngkuk Kim, Gil~Young Cho, and SungBin Lee.
\newblock Higher-order topological insulator in twisted bilayer graphene.
\newblock {\em Phys. Rev. Lett.}, 123:216803, Nov 2019.

\bibitem{TBGnew5}
Hiroki Isobe, Noah F.~Q. Yuan, and Liang Fu.
\newblock Unconventional superconductivity and density waves in twisted bilayer
  graphene.
\newblock {\em Phys. Rev. X}, 8:041041, Dec 2018.

\bibitem{TBGnew6}
Naoto Nakatsuji and Mikito Koshino.
\newblock Moir\'e disorder effect in twisted bilayer graphene.
\newblock {\em Phys. Rev. B}, 105:245408, Jun 2022.

\bibitem{TBGnew8}
Hoi~Chun Po, Liujun Zou, Ashvin Vishwanath, and T.~Senthil.
\newblock Origin of mott insulating behavior and superconductivity in twisted
  bilayer graphene.
\newblock {\em Phys. Rev. X}, 8:031089, Sep 2018.

\bibitem{TDBGnew1}
Narasimha~Raju Chebrolu, Bheema~Lingam Chittari, and Jeil Jung.
\newblock Flat bands in twisted double bilayer graphene.
\newblock {\em Phys. Rev. B}, 99:235417, Jun 2019.

\bibitem{TDBGnew2}
Xia Liang, Zachary A.~H. Goodwin, Valerio Vitale, Fabiano Corsetti, Arash~A.
  Mostofi, and Johannes Lischner.
\newblock Effect of bilayer stacking on the atomic and electronic structure of
  twisted double bilayer graphene.
\newblock {\em Phys. Rev. B}, 102:155146, Oct 2020.

\bibitem{TDBGnew3}
G.~William Burg, Jihang Zhu, Takashi Taniguchi, Kenji Watanabe, Allan~H.
  MacDonald, and Emanuel Tutuc.
\newblock Correlated insulating states in twisted double bilayer graphene.
\newblock {\em Phys. Rev. Lett.}, 123:197702, Nov 2019.

\bibitem{TDBGnew4}
Mikito Koshino.
\newblock Band structure and topological properties of twisted double bilayer
  graphene.
\newblock {\em Phys. Rev. B}, 99:235406, Jun 2019.

\bibitem{TDBGnew5}
J.~A. Crosse, Naoto Nakatsuji, Mikito Koshino, and Pilkyung Moon.
\newblock Hofstadter butterfly and the quantum hall effect in twisted double
  bilayer graphene.
\newblock {\em Phys. Rev. B}, 102:035421, Jul 2020.

\bibitem{TDBGnew6}
Yi-Xiang Wang, Fuxiang Li, and Zi-Yue Zhang.
\newblock Phase diagram and orbital chern insulator in twisted double bilayer
  graphene.
\newblock {\em Phys. Rev. B}, 103:115201, Mar 2021.

\bibitem{TBGnew7}
Yuncheng Mao, Daniele Guerci, and Christophe Mora.
\newblock Supermoir\'e low-energy effective theory of twisted trilayer
  graphene.
\newblock {\em Phys. Rev. B}, 107:125423, Mar 2023.

\bibitem{Calder}
M.~J. Calder\'on, A.~Camjayi, and E.~Bascones.
\newblock Mott correlations in abc graphene trilayer aligned with hbn.
\newblock {\em Phys. Rev. B}, 106:L081123, Aug 2022.

\bibitem{Chen_Signatures}
Guorui Chen, Aaron~L Sharpe, Patrick Gallagher, Ilan~T Rosen, Eli~J Fox, Lili
  Jiang, Bosai Lyu, Hongyuan Li, Kenji Watanabe, Takashi Taniguchi, Jeil Jung,
  Zhiwen Shi, David Goldhaber-Gordon, Yuanbo Zhang, and Feng Wang.
\newblock Signatures of tunable superconductivity in a trilayer graphene moire
  superlattice.
\newblock {\em Nature}, 572(7768):215--219, August 2019.

\bibitem{Chen2019}
Guorui Chen, Lili Jiang, Shuang Wu, Bosai Lyu, Hongyuan Li, Bheema~Lingam
  Chittari, Kenji Watanabe, Takashi Taniguchi, Zhiwen Shi, Jeil Jung, Yuanbo
  Zhang, and Feng Wang.
\newblock Evidence of a gate-tunable mott insulator in a trilayer graphene
  moire superlattice.
\newblock {\em Nature Physics}, 15:237--241, 2019.

\bibitem{Wang2020CorrelatedEP}
Lei Wang, En-Min Shih, Augusto Ghiotto, Lede Xian, Daniel~A. Rhodes, Cheng Tan,
  Martin Claassen, Dante~M. Kennes, Yusong Bai, Bumho Kim, Kenji Watanabe,
  Takashi Taniguchi, Xiaoyang Zhu, James Hone, Angel Rubio, Abhay~N. Pasupathy,
  and Cory~R. Dean.
\newblock Correlated electronic phases in twisted bilayer transition metal
  dichalcogenides.
\newblock {\em Nature Materials}, 19:861--866, 2020.

\bibitem{TMD1}
Mathieu B\'elanger, J\'er\^ome Fournier, and David S\'en\'echal.
\newblock Superconductivity in the twisted bilayer transition metal
  dichalcogenide ${\mathrm{wse}}_{2}$: A quantum cluster study.
\newblock {\em Phys. Rev. B}, 106:235135, Dec 2022.

\bibitem{TMD2}
Xueheng Kuang, Zhen Zhan, and Shengjun Yuan.
\newblock Flat-band plasmons in twisted bilayer transition metal
  dichalcogenides.
\newblock {\em Phys. Rev. B}, 105:245415, Jun 2022.

\bibitem{TMD3}
Mit~H. Naik and Manish Jain.
\newblock Ultraflatbands and shear solitons in moir\'e patterns of twisted
  bilayer transition metal dichalcogenides.
\newblock {\em Phys. Rev. Lett.}, 121:266401, Dec 2018.

\bibitem{TMD4}
Yi-Ming Wu, Zhengzhi Wu, and Hong Yao.
\newblock Pair-density-wave and chiral superconductivity in twisted bilayer
  transition metal dichalcogenides.
\newblock {\em Phys. Rev. Lett.}, 130:126001, Mar 2023.

\bibitem{TMD5}
Jiawei Zang, Jie Wang, Jennifer Cano, and Andrew~J. Millis.
\newblock Hartree-fock study of the moir\'e hubbard model for twisted bilayer
  transition metal dichalcogenides.
\newblock {\em Phys. Rev. B}, 104:075150, Aug 2021.

\bibitem{PhysRevB.82.035409}
Fan Zhang, Bhagawan Sahu, Hongki Min, and A.~H. MacDonald.
\newblock Band structure of $abc$-stacked graphene trilayers.
\newblock {\em Phys. Rev. B}, 82:035409, Jul 2010.

\bibitem{Anderson}
P.~W. Anderson.
\newblock Localized magnetic states in metals.
\newblock {\em Phys. Rev.}, 124:41--53, Oct 1961.

\bibitem{TriWarping}
Mikito Koshino and Edward McCann.
\newblock Trigonal warping and berry's phase $n\ensuremath{\pi}$ in abc-stacked
  multilayer graphene.
\newblock {\em Phys. Rev. B}, 80:165409, Oct 2009.

\bibitem{Uchoa}
Bruno Uchoa, Valeri~N. Kotov, N.~M.~R. Peres, and A.~H. Castro~Neto.
\newblock Localized magnetic states in graphene.
\newblock {\em Phys. Rev. Lett.}, 101(2):026805, Jul 2008.

\bibitem{Flensberg}
H.~Bruus and K.~Flensberg.
\newblock Many-body quantum theory in condensed matter physics, an
  introduction.
\newblock {\em (Oxford: Oxford University Press)}, 2012.

\bibitem{DFTCorrelation}
Bet\"ul Pamuk, Jacopo Baima, Francesco Mauri, and Matteo Calandra.
\newblock Magnetic gap opening in rhombohedral-stacked multilayer graphene from
  first principles.
\newblock {\em Phys. Rev. B}, 95:075422, Feb 2017.

\bibitem{AdatomAbove}
Kevin~T. Chan, J.~B. Neaton, and Marvin~L. Cohen.
\newblock First-principles study of metal adatom adsorption on graphene.
\newblock {\em Phys. Rev. B}, 77:235430, Jun 2008.

\bibitem{Criteria}
R~Van Pottelberge, D~Moldovan, S~P MilovanoviA, and F~M Peeters.
\newblock Molecular collapse in monolayer graphene.
\newblock {\em 2D Materials}, 6(4):045047, sep 2019.

\bibitem{Friedel}
C.~Dutreix and M.~I. Katsnelson.
\newblock Friedel oscillations at the surfaces of rhombohedral $n$-layer
  graphene.
\newblock {\em Phys. Rev. B}, 93:035413, Jan 2016.

\bibitem{TopoFano}
W.~C. Silva, W.~N. Mizobata, J.~E. Sanches, L.~S. Ricco, I.~A. Shelykh,
  M.~de~Souza, M.~S. Figueira, E.~Vernek, and A.~C. Seridonio.
\newblock Topological charge fano effect in multi-weyl semimetals.
\newblock {\em Phys. Rev. B}, 105:235135, Jun 2022.

\bibitem{Fano}
Andrey~E. Miroshnichenko, Sergej Flach, and Yuri~S. Kivshar.
\newblock Fano resonances in nanoscale structures.
\newblock {\em Rev. Mod. Phys.}, 82:2257--2298, Aug 2010.

\end{thebibliography}

\end{document}